\documentclass[11pt,a4paper]{article}

\usepackage[utf8]{inputenc}
\usepackage[T1]{fontenc}
\usepackage{geometry}
\usepackage{amsmath,amssymb,bm}
\usepackage{graphicx}
\usepackage{hyperref}
\usepackage{cite}

\hypersetup{
    colorlinks=true,
    linkcolor=blue,
    citecolor=blue,
    urlcolor=blue
}

\def\be{\begin{eqnarray}}
\def\ee{\end{eqnarray}}

\def\O{{\mathcal{O}}}
\def\calM{{\mathcal{M}}}
\def\mN{m_N}
\def\vr{{\vec r}}

\def\vp{{\vec p}}

\def\vs{{\vec \sigma}}

\def\vA{{\vec A}}
\def\no{\nonumber}

\begin{document}

\title{\textbf{The chiral filter}}

\author{
  Tae-Sun Park\thanks{Email: \href{mailto:tspark@ibs.re.kr}{tspark@ibs.re.kr}} \\[0.5em]
  \small Center for Exotic Nuclear Studies (CENS), Institute for Basic Science, \\
  \small Daejeon 34126, Republic of Korea
}

\date{\today}

\maketitle

\begin{abstract}
Thermal neutron capture on a proton, $np\to d\gamma$, is measured to a fraction of a percent; the impulse approximation misses nine percent of it. This article follows that gap through five decades of an idea of Mannque Rho's, the chiral filter. Chiral symmetry fixes the pion-exchange current that fills the gap, and power counting explains why nothing else competes. The missing nine percent becomes a prediction, correct within errors. On the unprotected side, one constant fixed by tritium beta decay predicted solar $pp$ fusion and the {\it hep} process; $g_A^*$ remains open.
\vspace{1em}

\noindent\textbf{Keywords:} Chiral filter, Soft pion, Chiral perturbation theory, Meson-exchange currents, $np\to d\gamma$, Solar nuclear fusion
\end{abstract}

\section{Introduction: The missing nine percent}\label{Xsec1-1}

Consider the simplest radiative process in nuclear physics, the capture of
a thermal neutron by a proton,
\be
n+p &\longrightarrow& d+\gamma .
\ee
The cross section has long been measured to $0.15\%$,
\be
\sigma_{\rm exp} &=& 334.2\pm0.5\ {\rm mb}.
\ee
The transition is dominantly isovector magnetic dipole (M1). The impulse
approximation, in which the photon couples to one nucleon at a time, gives
\cite{Rho1997}
\be
\sigma_{\rm imp} &=& 305.6\ {\rm mb}.
\ee
About nine percent of the cross section is missing. The missing part is
due to two-body currents,
\be
J^\mu \;=\; J^\mu_{\rm 1B}+J^\mu_{\rm 2B}+\cdots .
\ee
The question is whether the missing nine percent can be {\it predicted},
that is, whether the two-body current is fixed by something better than a
model.

This article follows that gap across five decades, because the answer is
an idea of Mannque Rho's: the {\it chiral filter}. The
idea did not arrive in finished form. It was an observation (1978), then a
hypothesis (1981), then a power-counting statement (1991), and finally a
verified prediction (1995). At the end I describe, more briefly, what the same idea did
in the channels it does not protect. That other side reaches from the
axial charge through the solar fusion reactions to the problem of $g_A^*$,
which occupied him to the end of his life.

The story is narrower than a review of his work, but it captures something
characteristic of the way he did physics. He often identified the physical
pattern first and only later found the language in which the pattern could
be made systematic.

\section{1972: One-pion exchange fills the gap}\label{Xsec2-2}

That exchange currents exist was never in doubt. The problem, before
effective field theory, was that there were too many of them
\cite{Riska1989}. Besides one-pion exchange one could consider $\rho$ and
$\omega$ exchange, $\Delta(1232)$ excitation, multi-pion exchange,
relativistic corrections, form factors, and various off-shell
prescriptions:
\be
J^\mu_{\rm 2B} &=& J^\mu_{\pi}+J^\mu_{\rho}+J^\mu_{\omega}
 +J^\mu_{\Delta}+J^\mu_{2\pi}+\cdots .
\ee
Each term could be calculated, often with great effort. But there was no
expansion parameter that told us which terms are needed at a given
accuracy. The labels are not even unique: different field representations
move strength between diagrams, and large short-range terms tended to
cancel among themselves. Gauge invariance constrains the electric
multipoles strongly, through Siegert's theorem~\cite{Siegert1937,Riska1989},
but the magnetic and axial channels have no such protection. Our
reaction, an isovector M1 transition, therefore sits where the
exchange currents are most model dependent.

Against this background, Riska and Brown \cite{RiskaBrown1972} found that
most of the nine-percent gap in $np\to d\gamma$ is removed by the
one-pion-exchange two-body currents alone. The shorter-range
mechanisms played only a secondary role. But in the language of the time
one could not yet say {\it why} this was trustworthy, that is, why the one
mechanism that worked should be believed over the many that could be added.

Part of the reason was already available. Chemtob and Rho
\cite{ChemtobRho1971} had combined the standard treatment of two-body
currents with the low-energy theorems of PCAC and current algebra. Among
the exchanged mesons, chiral symmetry constrains the pion and only the
pion. In the soft limit, an amplitude with one extra pion
reduces to a commutator of the axial charge with the relevant operator
\cite{AdlerDashen,ChemtobRho1971},
\be
\calM(X\rightarrow Y+\pi^a(q\simeq0))
&\sim& {i\over f_\pi}
\langle Y|[Q_5^a,\,\O_X]|X\rangle ,
\ee
up to pole terms and corrections that vanish with the pion four-momentum.
Here $\O_X$ is the operator responsible for the transition $X\to Y$; in our
case, the electroweak current. The leading soft-pion exchange current is
therefore fixed by symmetry, independent of the model for the nuclear
short-distance dynamics. The one mechanism that filled the gap was also
the one mechanism that symmetry could pin down. This coincidence is the
starting point of the chiral filter.

\section{1978: The soft-pion classification}\label{Xsec3-3}

Kubodera, Delorme and Rho \cite{KDR1978} turned the coincidence into a
classification. They examined the four components of the electroweak
current, the time and space components of the vector and axial currents,
according to the fate of the soft-pion exchange. The result fits in one table:
\begin{equation}
\begin{array}{llll}
\text{current} & \text{one-body} & \text{soft-}\pi\, \text{two-body}&  \\
A^0\ \text{(axial charge)} & \text{suppressed}\ (\propto 1/\mN) &
\text{unsuppressed} & \text{enhanced}\\
{{\vec V}}\ \text{(isovector M1)} & \text{suppressed}\ (\propto 1/\mN) &
\text{unsuppressed} & \text{enhanced}\\
\vA\ \text{(Gamow--Teller)} & \O(1) & \text{suppressed} &
\text{filtered out}\\
V^0\ \text{(charge)} & \O(1) & \text{suppressed} &
\text{filtered out}
\end{array}
\end{equation}
Our reaction sits in the second row. The one-body M1 operator carries a
$1/\mN$ suppression and the soft-pion two-body operator does not, so the
two-body term is unusually visible, and its structure is fixed by symmetry.
In the last two rows the soft pion itself is suppressed. In Rho's
later words \cite{Rho1991}, the nucleus ``played a role of `filtering' out
all but the longest-wavelength component, namely, the soft-pion-exchange
term, in response to long-wavelength weak and electromagnetic probes.''

Two clarifications are needed to state the claim accurately. First, even
in the protected rows the one-body operator dominates the matrix element;
in $np\to d\gamma$ the impulse approximation gives about $95\%$ of the
amplitude. What the $1/\mN$ suppression does is make the two-body
correction much larger than one would naively expect, and, more
importantly, calculable. Second, at this stage the statement concerned
soft-pion kinematics: it is exact only in the limit of vanishing pion
four-momentum. The pion exchanged between two nucleons carries a momentum
of the order of the pion mass or the typical nucleon momentum, small on
the chiral scale but not zero, and the theorem by itself says nothing
about the accuracy of the extrapolation.
The nontrivial finding was that the soft-pion result remains a
good approximation at such momenta, because the pion is the lightest
hadron and the longest-ranged carrier of the nuclear
interaction~\cite{Riska1989}: its propagator carries the long-distance part
of the operator, while everything heavier is localized at short distances.

\section{1981: The chiral filter hypothesis}\label{Xsec4-4}

What happened next was, in Rho's words, ``a complete surprise'':
the soft-pion predictions ``worked quantitatively well to large momentum
transfer, of the order of GeV'' \cite{Rho1991}. The M1 operator of our
reaction, probed at high momentum transfer in deuteron threshold
electrodisintegration, and its analogues in the trinucleon magnetic form
factors, kept following the soft-pion-dominated prediction deep into a
region where the derivation gave it no right to work.

The scales make the surprise concrete. The soft-pion operator has one-pion
range, $\sim\!1/m_\pi\simeq1.4$ fm. The competing mechanisms, vector
mesons and baryon resonances, live at $1/m_\rho\simeq0.25$ fm and below.
At momentum transfers of order a GeV the probe resolves distances where the
heavy mesons should dominate. Either the successes were an accident, or
the short-distance mechanisms were being systematically suppressed wherever
the soft pion was present.

This success gave the observation its name: the {\it chiral filter
hypothesis} \cite{RhoBrown1981,Rho1982}. Whenever the soft pion is
allowed, it dominates, and everything else stays small. At this point
there was no derivation. It was a pattern in the data, and for a decade it
stood as an open question: why should the short-distance mechanisms stay
silent exactly where the soft pion is present? Posing that question
sharply required a framework that did not yet exist.

\section{1991: The power-counting explanation}\label{Xsec5-5}

The framework came from Weinberg's formulation of nuclear forces in chiral
perturbation theory \cite{Weinberg1990,Weinberg1991,Epelbaum2009},
following his ``folk theorem'' for effective field theory
\cite{Weinberg1979}: identify the low-energy degrees of freedom, write the
most general Lagrangian consistent with the symmetries, and order the
operators by a power counting. A heavy meson no longer needs to be kept
explicitly; its low-momentum effect is expanded in local operators and
absorbed into low-energy constants. The old question, ``which
meson-exchange graphs should we trust?'', becomes an operational one:
``which operators occur at a given chiral order?''

The timing of this development was also when I first met Mannque Rho. In
1990, while I was a beginning doctoral student at Seoul National
University, he came to give a special series of lectures at the invitation
of Professor Dong-Pil Min. At the first lecture he asked anyone who did
not know chiral symmetry to raise a hand. I knew pieces of the subject
but did not feel that I could say that I ``knew'' it, so I raised mine; I
was the only one. He then asked the audience which nucleus is the most
stable, and I answered
$^4$He. He laughed and said that I had been right that I did not know chiral
symmetry. The answer he wanted was iron, the nucleus with the lowest
energy per nucleon. Later in the same series he said that anyone who
already knew what he was explaining could leave. The material at that
moment seemed elementary to me, so I left; again I was the only one. It
was an awkward introduction, but the
historical timing was perfect. Weinberg's formulation had just appeared,
and he soon asked me to work out what it implied for nuclear exchange
currents. That problem became my thesis.

In his 1991 Letter \cite{Rho1991}, Rho applied Weinberg's counting to
the currents and derived the filter. In the form he later used to teach it
\cite{Rho1997}, a physical amplitude with $E_N$ external nucleon lines
scales as $A\sim Q^\nu F(Q/\Lambda_\chi)$, with
\be
\nu &=& 4-2C+2L-\left({E_N\over2}+E_{\rm ext}\right)+\sum_i \bar\nu_i ,
\ee
where $C$ counts clusters, $L$ loops, $E_{\rm ext}$ external fields, and
the sum runs over the vertices, each carrying the index
\be
\bar\nu_i &=& d_i+{n_i\over2}+e_i-2 ,
\ee
with $d_i$ derivatives (or pion-mass insertions), $n_i$ nucleon fields and
$e_i$ external fields at the vertex. Chiral symmetry guarantees
\be
\bar\nu_i &\geq& 0 .
\ee

The filter now follows from two short computations. In the nuclear force,
the pion--nucleon vertex has $d_i=1$, $n_i=2$, so $\bar\nu_i=0$; the
derivative-free four-Fermi contact term has $d_i=0$, $n_i=4$, so again
$\bar\nu_i=0$. The longest-range and shortest-range parts of the force
enter on the same footing:
\be
V_{\rm LO} &\sim& V_{1\pi}+V_{\rm contact}.
\ee
Now attach a slowly varying external field. Attached to the pion-exchange
term, $e_i+d_i=1$ keeps $\bar\nu_i=0$: the soft-pion current stays at
leading order. Attached to the contact term, one needs $e_i=1$ on top of $n_i/2=2$, so
$\bar\nu_i\ge1$: the short-range current is pushed down to higher order,
\be
J^{\mu}_{\rm 2B,LO} &\sim& J^{\mu}_{1\pi,\,\rm soft}.
\ee
This is
``the first solid justification for the notion of
the chiral filter hypothesis'' \cite{Rho1991}.
The physical reason fits in
one sentence:
the leading heavy-meson exchanges
are taken into account by
a derivative-free contact term
which does not couple to the
external field.
Interestingly, as discussed in Ref.~\cite{KDR1978}, the one-body operator
of the channels where the soft-pion exchange survives is suppressed by
$1/\mN$. Since a two-body current is kinematically suppressed by
$(Q/\Lambda_\chi){}2$, the soft one-pion exchange then appears at order
$Q/\Lambda_\chi$, next-to-leading order (NLO), relative to the one-body
contribution. Loop corrections and, in these channels, the short-range counterterms enter
only at $(Q/\Lambda_\chi){}3$, N$^3$LO, which leaves the soft one-pion
exchange clean and visible.

The mystery of 1981 was thereby reduced to counting. In the M1 and
axial-charge channels the symmetry-fixed one-pion-exchange current is the
first correction to the one-body term, and the next corrections, loops and
counterterms, enter only two orders later.
In the Gamow-Teller
(and vector charge) channels the
one-body operator is unsuppressed while the soft pion needs one
next-to-leading-order vertex, so the two-body current starts three orders
down, together with contact operators: in the Gamow--Teller channel the
contact term carries $\bar\nu_i=1$ and arrives at the same order as the
pion. Note that the
short-range current sits at N$^3$LO relative to the one-body term in the
M1 and Gamow--Teller channels alike. What distinguishes the protected
channels is not the order of the short-range physics but the position of
the pion: two orders ahead of it in M1, at the same order in
Gamow--Teller. Protection in two channels, none in
the other two: exactly the table of 1978, now derived.

Rho also stated clearly what the filter costs \cite{Rho1997}:
\begin{quote}
``This `chiral filtering' is both a good news and a bad news. It is a good
news in that meson-exchange currents can be under control with the
dominance of Goldstone pions without interference from poorly-understood
short-range degrees of freedom. It is a bad news since the pion dominance
means that, unless accidentally suppressed, pions will not allow us to
learn short-distance physics through exchange currents.''
\end{quote}

\section{1995: The gap becomes a prediction}\label{Xsec6-6}

One complication is special to nuclear systems. Reducible diagrams are
enhanced by small energy
denominators~\cite{Weinberg1990,Weinberg1991,Epelbaum2009} and must be
iterated to all orders to produce bound and scattering states. Weinberg's
prescription is to build an irreducible kernel by chiral counting and
evaluate it between nonperturbative wave functions,
\be
\calM &=& \langle \Psi_f|\,(J_{\rm 1B}+J_{\rm 2B}+\cdots)\,|\Psi_i\rangle .
\ee
This separation between operator and wave function became the practical
basis of our work: the wave functions from the best available
phenomenology, the operator from the chiral expansion, order by order.
The formal mismatch between the wave functions and the operator is
discussed in the next section.

Our one-loop calculation \cite{ParkMinRho1995} carried the expansion to
one-loop order in the current (one-body $\nu=-2$, tree two-body $\nu=-1$,
one-loop $\nu=+1$; the order $\nu=0$ vanishes, so the loop is an N$^3$LO
correction) and gave the exchange-current correction
\be
\delta_{\rm MEC} &=& (4.5\pm0.3)\%
\ee
in the amplitude, hence
\be
\sigma_{\rm th} &=& 334\pm2\ {\rm mb},
\ee
against $\sigma_{{\exp}}=334.2\pm0.5$ mb. The missing nine percent of the
cross section was no longer a phenomenological patch. It was a prediction.

\begin{figure}[t]
\centering
\includegraphics[width=0.6\linewidth]{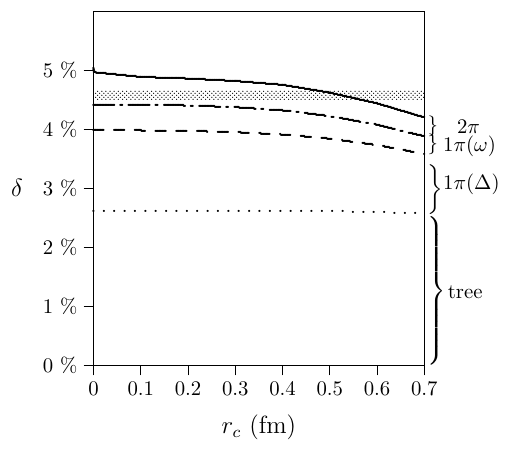}
\caption{Exchange-current correction $\delta_{\rm MEC}$ to the
$np\to d\gamma$ amplitude as a function of the short-distance cutoff
radius $r_c$. The curves show the result as the soft-pion tree,
$1\pi(\Delta)$, $1\pi(\omega)$, and genuine two-pion-exchange
contributions are successively included. 
The horizontal band indicates the value required by experiment.
Adapted from Ref.~\cite{ParkMinRho1995}.}
\label{fig:npcap-amplitude}
\end{figure}

As shown in Fig.~\ref{fig:npcap-amplitude}, 
more revealing than the final number was the hierarchy of the
contributions. The generalized-tree contribution, comprising the
soft-pion tree term together with the $1\pi(\omega)$ and
$1\pi(\Delta)$ terms, dominated. The genuine two-pion-exchange loops
stayed below one percent of the amplitude. The diagrams could be
compared one by one with the mechanisms of the older MEC literature,
but they no longer entered as an unranked catalogue: the resonance
content of the old language reappeared through the finite counterterms
of the $\pi VNN$-type vertices, saturated at low photon energy by the
$\omega$- and $\Delta$-exchange tree graphs, with an assigned order and
no independent freedom. Going to one loop did not reveal a hidden large
mechanism. It quantified the corrections to the soft-pion picture.
Twenty-three years after Riska and Brown, the one-pion-exchange origin
of the missing cross section was no longer an empirical observation.
It was a consequence of the counting, checked at one loop.

There is, however, a caveat. The current up to N$^3$LO contains a
short-range counterterm contribution with an unknown coefficient, a
low-energy constant not fixed by chiral symmetry. The constant can be
determined from an independent observable, for example the magnetic moments
of $^3$H and $^3$He, but that requires accurate three-body wave functions.
At the time our machinery was limited to the two-nucleon sector, so the
counterterm contribution was set to zero. 
The price is also visible in Fig.~\ref{fig:npcap-amplitude},
adapted from Fig.~2 of Ref.~\cite{ParkMinRho1995}:
the result retains a sizable dependence on
the momentum cutoff, and it is this dependence that sets the quoted error
of $\pm2$ mb, four times the experimental one. This was not a mere
numerical nuisance. Residual cutoff dependence has a meaning: it measures
how much of the required short-distance information is still missing. An
EFT does not remove short-distance physics; it tells us how much of it is
needed, packs it into a finite number of constants, and points to the
observables that can fix them. This point set the direction of the later
work.

\section{The same counting in other channels}\label{Xsec7-7}

If the filter is right, the same hierarchy must appear wherever the table
of 1978 says it should, and fail in a specific and organized way where the
table says it should not. Both happened.

\subsection{The axial charge: The same protection, magnified}\label{Xsec8-7.1}

The first row of the table is the axial charge, where the one-body operator
is relativistically suppressed,
\be
A^0_{\rm 1B} \;\sim\; g_A\,{\vs\cdot(\vp'+\vp)\over 2\mN}\,\tau^a ,
\qquad
A^0_{\rm 2B} \;\simeq\; A^0_{{\rm soft}\mbox{-}\pi},
\ee
so the protected two-body term is not a ten-percent correction but tens of
percent. Warburton's analysis of first-forbidden beta decays
\cite{Warburton1991,Warburton1991C} found
$\epsilon_{\rm MEC}=M_{{\exp}}/M_{\rm sp}\simeq1.5$ in light nuclei,
rising to $1.8$--$2.0$ in the lead region \cite{Rho1997}. Our one-loop
analysis of the operator \cite{ParkMinRhoAxial}, applied with shell-model
wave functions from $A=16$ to $A=208$ \cite{ParkTownerKubodera1994}, found
the loop contribution to be $8$--$10\%$ of the soft-pion tree term. This
is of order $(Q/\Lambda_\chi){}2$, right where the counting puts it, and it
was nearly independent of the nucleus and of the particular transition.
The correction is large but controlled, the same pattern as in
$np\to d\gamma$. Together, the two channels confirmed the hypothesis
quantitatively.

\subsection{The unprotected side: renormalization and one constant}\label{Xsec9-7.2}

The Gamow--Teller channel is the row without protection: the soft pion
arrives three orders down, together with a contact operator
\cite{ParkEtAl2003,KuboderaPark2004,Baroni2016}. One might think this is
the end of predictive power. It was instead the start of a second program,
because the counting still organizes what it cannot protect. To the order
relevant for precision low-energy weak processes, the unknown
short-distance physics of the axial two-body current is a single constant
$\hat d^R$:
\be
\vA_{\rm 2B}
&=& \vA_{\rm 2B}^{\pi}
-
\frac{2 g_A}{\mN f_\pi^2}
\hat d^R
\sum_{i<j}(\tau_i \vec\sigma_i
+ \tau_j \vec\sigma_j)
\delta^{(3)}(\vec r_{ij})
+\cdots .
\ee
When symmetry is not enough, the EFT does not fail silently. It tells you
exactly what to measure.

The procedure rests on a renormalization argument. The problematic physics
is short ranged, and in EFT everything short ranged is described by local
operators,
\be
\O_{\rm short} &=& \sum_n c_n\,\nabla^{2n}\delta^{(3)}(\vr)
\;=\; c_0\,\delta^{(3)}(\vr)+\cdots .
\ee
We impose renormalization conditions on the coefficients $c_n$: they are
adjusted to reproduce known experimental data. In doing so they absorb the
mismatch, the model dependence, and the residual inconsistency of the
calculational scheme. The values of the $c_n$ are model dependent and
regulator dependent; they are not observables. The renormalized matrix
elements $\langle\Psi_f|\O|\Psi_i\rangle$ are model independent,
provided that
the wave functions have the correct long-range behavior.
This is
the content of the hybrid scheme we called EFT$^*$
\cite{ParkEtAl2003,KuboderaPark2004}: chiral transition operators,
renormalization conditions imposed on their contact terms, and accurate
wave functions.

For the axial current the datum is tritium beta decay, whose rate is known
accurately and computable with high-quality three-body wave functions. For
each regulator $\Lambda$ one fits $\hat d^R(\Lambda)$ to the tritium rate
and predicts everything else with no adjustable parameter. The
renormalization is visible in the numbers. With a Gaussian regulator of
scale $\Lambda=500$, $600$, $800$ MeV, the solar $pp$ fusion matrix
elements behave as
\be
\Lambda=500:&& \hat d^R=1.00,\quad \langle{\rm 1B}\rangle=4.85,\quad
\langle{\rm 2B}\rangle=0.076-0.035\,\hat d^R=0.041,\no\\
\Lambda=600:&& \hat d^R=1.78,\quad \langle{\rm 1B}\rangle=4.85,\quad
\langle{\rm 2B}\rangle=0.097-0.031\,\hat d^R=0.042,\no\\
\Lambda=800:&& \hat d^R=3.90,\quad \langle{\rm 1B}\rangle=4.85,\quad
\langle{\rm 2B}\rangle=0.129-0.022\,\hat d^R=0.042.
\no\\
&&
\ee
The fitted $\hat d^R$ varies by a factor of four; the bare two-body pieces
by almost a factor of two. Neither is observable, and their combination
does not move. The result,
\be
S_{pp}(0) &=& 3.94\,\left(1\pm0.15\%\pm0.10\%\right)
\times10^{-25}\ {\rm MeV\,b},
\ee
is a $0.2\%$ determination of the reaction that sets the solar timescale
\cite{Park1998,ParkEtAl2003}. It was an answer to John Bahcall, whose
request for a $pp$ $S$ factor with a defensible error bar had started our
involvement, in the years when Super-Kamiokande and SNO were turning solar
neutrinos into a precision laboratory.

The same current, with the same $\hat d^R$ and nothing new to tune, was
then applied to the so-called
{\it hep} process,
${}^3{\rm He}+p\to{}^4{\rm He}+e^+\nu_e$, a reaction whose published
$S$-factor estimates had varied over a factor of $500$ since 1952
\cite{KuboderaPark2004}. The reason is a symmetry accident. The leading
one-body matrix element is strongly suppressed
due to pseudo-orthogonality between
the initial and the final wave functions,
\be
\left\langle {}^4{\rm He}\left|
g_A \sum_i \vs_i\tau_i
\right|p+{}^3{\rm He}\right\rangle &\simeq& 0 ,
\ee
and the one- and two-body terms cancel to a third of their size
\cite{Marcucci2000,Marcucci2001}. We obtained
\be
S_{\rm hep}(0) &=& (8.6\pm1.3)\times10^{-20}\ {\rm keV\,b},
\ee
where the error spans the residual cutoff dependence, amplified by the
cancellation \cite{ParkEtAl2003}. The uncertainty is much larger than for
$pp$: an observable can be correctly renormalized and still be hard to
predict when its leading amplitude is accidentally small. But the number
should be read against its history. A quantity whose published estimates
had wandered over a factor of $500$ was pinned down to $15\%$, with its
short-distance uncertainty tied to a measured few-body observable. The
timing gave the number weight. The {\it hep} neutrinos are the most
energetic solar
neutrinos, reaching beyond the endpoint of the $^8$B spectrum, and in
those years an apparent excess in the highest-energy bins of
Super-Kamiokande could be attributed to an enhanced {\it hep} flux;
testing that interpretation required a {\it hep} $S$ factor with a
defensible error bar,
which had never existed before \cite{KuboderaPark2004}. The result was
adopted in the community evaluation of solar fusion cross sections
\cite{SolarFusionII}.

The logic behind both predictions is short: tritium beta decay fixes
$\hat d^R(\Lambda)$, and the same $\hat d^R(\Lambda)$ then predicts $pp$
and {\it hep}. One low-energy constant is transported among nuclei of
different
structure, which is a stronger test than agreement with any single
observable. The filter did not divide processes into good and bad ones.
It told us where symmetry was enough and where an additional datum was
required.

\subsection{The spatial Gamow--Teller channel}\label{Xsec10-7.3}

The spatial Gamow--Teller operator itself is the hardest case of the
unprotected side. There is no soft-pion protection, and short-range
correlations and many-body dynamics enter early. Rho had been
concerned with the associated ``quenching'' of the effective axial coupling
long before the chiral filter acquired its name, and connected the problem
successively to pion--nucleus dynamics, $\Delta$--hole correlations,
in-medium hadron properties, and finally to Fermi-liquid and scale-symmetry
ideas \cite{Rho1974,OsetRho1979,BrownRho1991,LiMaRho2018,MaRho2020}. The
interpretation changed over the years, but the question did not: why does
nuclear Gamow--Teller phenomenology so often look as if
$g_A^{\rm eff}\simeq1$ rather than the free-space value $g_A\simeq1.27$
\cite{Towner1987}? A large part of the answer is already stated,
explicitly, in the 1991 Letter \cite{Rho1991}: in this unprotected channel
the quenching is attributed to short-range dynamics, a zero-range
$\Delta$--hole interaction, in the language of the counting an $n_i=4$
contact term, acting on top of the correlations carried by the wave
functions. Modern ab initio calculations with chiral interactions and
currents arrive at essentially this picture: with accurate wave functions
and consistent two-body currents, most of the quenching is accounted
for~\cite{Baroni2016,Gysbers2019}. What he wanted was more than this
accounting. The same passage of the Letter reads the quenching as
suggesting ``a change of chiral-symmetry scale in dense medium,'' and that
is the thread he followed afterwards: not whether good wave functions can
reproduce the fitted $g_A^{\rm eff}$, but whether an intrinsic in-medium
constant $g_A^*$ can be understood directly, as a property of how QCD
symmetries are realized in dense matter. The unprotected side of the filter is not empty; it marks
where the nontrivial nuclear dynamics must be confronted. I leave that
story for another occasion.

\section{Epilogue}\label{Xsec11-8}

The missing nine percent with which this story began was filled in 1972.
What first appeared as a correction was singled out in 1978, became a
hypothesis in 1981, found its explanation in chiral power counting in
1991, and by 1995 had become a quantitative prediction: $334\pm2$ mb,
compared with the measured $334.2\pm0.5$ mb. The same counting was later
carried to the unprotected side of the filter, from solar fusion to the
long-standing problem of $g_A^*$.

Looking back over this history, I am struck by how little the central
question changed. The early current-algebra work and modern EFT ask
essentially the same thing in different languages: which part of a
nuclear amplitude is fixed by symmetry, which part represents unresolved
short-distance physics, and at what order does each enter? The chiral
filter sits between the two eras. Today it has been embodied in the
formalism: modern chiral current operators
\cite{Pastore2009,Baroni2016} reproduce the same leading hierarchy
directly from power counting. In this sense, what began as the
chiral-filter hypothesis has become a consequence of the EFT power
counting.

It seems to me that, for Mannque, a physical picture and a systematic
formalism were not alternatives. The soft-pion idea acquired a deeper
meaning when Weinberg's counting showed why it worked and where its
limitations lay. The chiral filter is therefore not only a piece of
nuclear-current phenomenology. It is an example of a style of doing
physics in which physical insight comes first, and a systematic framework
is then asked to sharpen, test, and sometimes correct it. The filter is
where physical insight and systematic expansion meet. To me, that is very
much Mannque.

I close with a personal debt. Almost all of the physics recalled in this article I learned from him. He answered every question I asked with great sincerity. In those years my daily routine was to read his e-mails, think about his ideas, and look for answers to his questions. I also remember that, whenever I made English corrections during revisions of our drafts, he almost always kept my version to the end, even when I suspected that my changes were not quite to his taste. When I once brought him a question about life rather than physics, he carried it with him and came back to it long afterwards. 

When my position was uncertain, he worried about my future as if it were
his own concern; my first postdoctoral position came through him. He cared
for his students with a closeness that went well beyond physics. Thank you,
Mannque.

\section*{Declaration of competing interest}
The authors declare that they have no known competing financial
interests or personal relationships that could have appeared to
influence the work reported in this paper.

\section*{Acknowledgments}
I am grateful to the many collaborators with whom the work recalled here was carried out, in particular Professors Dong-Pil Min, Kuniharu Kubodera, and I. S. Towner, and the collaborators in the later few-body calculations. Above all, I remember Mannque Rho for the questions he asked, the physical pictures he insisted on finding, and the directions those questions opened for us.
This work was supported by the Institute for Basic Science (IBS-R031-D1).

\section*{Data availability}
No data was used for the research described in the article.

\end{document}